\documentclass[%
 reprint,
 amsmath,amssymb,
aps,
pra,
showkeys
]{revtex4-2}

\usepackage{siunitx}
\usepackage{graphicx}% Include figure files
\usepackage{dcolumn}% Align table columns on decimal point
\usepackage{bm}% bold math
\usepackage{amsmath}
\usepackage{tikz}
\usepackage{xspace}
\usetikzlibrary{matrix,fit,calc} 
\usepackage[colorlinks,citecolor=red,urlcolor=blue,bookmarks=false,hypertexnames=true]{hyperref}
\usepackage{multirow}

\newcommand{\eqab}{Eq.~}
\newcommand{\figab}{Fig.~}
\newcommand{\tabab}{Tab.~}
\newcommand{\catwop}{Ca$^{2+}$\xspace}

\begin{document}

%\preprint{APS/123-QED}

\title{Theoretical Study of Classical Dynamics of  Electron in  Two-Frequency Paul Trap}
% Force line breaks with \\

\author{Payman Mahmoudi}
\email{payman.mahmoudi@jh-inst.cas.cz}
\affiliation{Heyrovský Institute of the CAS, Dolejškova 2155/3, 182 23 Prague 8, Czech Republic}
\author{Michal Tarana}
\email{michal.tarana@jh-inst.cas.cz}
\affiliation{Heyrovský Institute of the CAS, Dolejškova 2155/3, 182 23 Prague 8, Czech Republic}
\author{Michal Hejduk}
\email{michal.hejduk@matfyz.cuni.cz}
\affiliation{Charles University, Faculty of Mathematics and Physics, V~Holešovičkách 2, 180 00 Praha 8, Czech Republic}
\begin{abstract}
We present a theoretical study of trapping an electron in a two-frequency Paul trap, motivated by future hybrid architecture consisting of a doubly charged calcium ion (Ca$^{2+}$) and an electron. By isolating the electron from ionic-electron interactions, we formulate the equations of motion for the electron in the time-dependent trap potentials and cast them in the Hill form. Stability and dynamics are then analysed in terms of the dimensionless q-parameters, which encode the applied voltages, drive frequencies, trap geometry, and the particle's charge-to-mass ratio. Under the combined action of the two radio-frequency fields, the trapped electron's classical motion is shown to be approximated well by a slow secular oscillation modulated by two distinct motion sidebands originating from the low- and high-frequency RF drives. The approximation method presented here enables a quick evaluation of the relation between the device-design q-parameters and the electron motion. It offers a crucial foundation for developing physical models of electron–ion interactions in electron-ion traps.
\end{abstract}

\keywords{}%Use showkeys class option if keyword
                              %display desired
\maketitle

%\tableofcontents

\section{Introduction}
The controlled trapping of single electrons has emerged as a promising direction in quantum science and technology. Owing to their extremely low mass and strong coupling to electromagnetic fields, trapped electrons are attractive not only for quantum computing but also for high-precision sensing \cite{Gabrielse2023} and quantum simulation \cite{Hensgens2017}.

In the context of quantum computing, recent experimental progress has demonstrated the feasibility of using trapped electrons as qubits~\cite{Yu2022}. Electron trapping using Paul traps has been explored in room-temperature microwave trap architectures~\cite{Matthiesen20}, where coherent control and readout of electron quantum states are actively pursued. A parallel line of research focuses on electrons trapped in vacuum above the surface of superfluid helium, a system that enables exceptionally clean environments and long coherence times~\cite{Platzman1999, Dykman2003, Dahm2003, Schuster2010}. These electrons have been proposed for use as both charge qubits~\cite{Kawakami2023, Shi2014} and spin qubits~\cite{Lyon2006, Dykman2023}, and several experimental milestones in trapping, detection, and coupling to superconducting circuits have been achieved~\cite{Rousseau2007, Koolstra2019}. Furthermore, recent work has shown that electrons trapped above solidified noble gases—particularly solid neon—can function as natural charge qubits, with experimental demonstrations of long coherence times and strong coupling to quantum hardware~\cite{Zhou2022, Zhou2023}.

Beyond free electrons in vacuum, similar goals have been pursued in condensed matter systems, such as quantum dots~\cite{Yoneda2018} and donor-bound electrons in silicon~\cite{Laucht2017}, and in atomic ion systems where valence electrons are confined. However, these differ fundamentally from the goal of trapping and manipulating free, isolated electrons in vacuum.

Hybrid quantum systems combining trapped electrons with superconducting electronics have also been theoretically explored~\cite{Daniilidis2013, Kotler2017}, and spin readout schemes for trapped-electron qubits have been proposed~\cite{Peng2017}. Motivated by these developments, we propose and theoretically investigate a novel hybrid quantum system in which an electron and a doubly charged calcium ion are simultaneously trapped in a two-frequency Paul trap. The combination of light and heavy particles in a single trap offers a promising route towards new hybrid quantum architectures.

Historically, trapping electrons has been achieved using Penning traps \cite{Malmberg1980,Driscoll1983}, which employ strong magnetic and electrostatic fields to confine charged particles. More recently, efforts have been directed toward trapping electrons in radio-frequency (RF) traps, such as Paul traps \cite{Matthiesen20,Yu2022}, which are widely used for  trapping ions \cite{ghosh2007ion,Leibfried}. While these works demonstrate the feasibility of electron or ion trapping separately, achieving simultaneous confinement of such vastly different species in the same Paul trap presents a fundamental challenge. The RF fields that stabilize a heavy ion are not suitable to stably confine a significantly lighter and more mobile electron that tends to escape.

To simultaneously trap two different species of charged particles, two-frequency Paul traps have been proposed and implemented to independently tune the trapping conditions for each species \cite{Foot2018}. These traps operate by simultaneously applying multiple RF voltages with different frequencies and amplitudes to the same electrodes, thereby generating a superposition of effective pseudopotentials that can be engineered to stably trap both light and heavy charged species. The motivation for co-trapping particles with different mass-to-charge ratios stems from a range of applications, e.g. sympathetic cooling of heavy particles by collisions with light ones \cite{Miao2023}, or quantum logic with mixed-species systems \cite{Barrett2023}.

From a theoretical point of view, the dynamics of particles in a two-frequency RF field are governed by a  set of  Hill differential equations \cite{McLachlan1947}. While the traditional single-frequency Paul trap exhibits well-known stability maps \cite{paul1958, ghosh2007ion, Raymond1995}, the introduction of a second frequency creates a far more intricate structure in phase space, giving rise to new regions of parametric stability and instability. Several theoretical and experimental studies have extended the classical pseudopotential approximation to handle multi-frequency trap fields. \citet{Trypogeorgos2016} developed stability frameworks and simulations of two-frequency Paul traps. \citet{Foot2018} derived analytic stability conditions, including damping effects. Similarly, \citet{Possa2016} explored two-frequency stability diagrams and demonstrated that adjusting the frequency and amplitude ratios improves control over trapping.

In this work, we present a theoretical study of trapping an electron in a two-frequency Paul trap, motivated by future hybrid architecture consisting of a calcium ion ($\text{Ca}^{2+}$) and an electron. The interaction of the electron with the ion in the trap adds additional level of complexity to the system. If the electron is weakly bound by the ion in a Rydberg state, the additional time-dependent trapping RF fields can affect it or even detach the electron from the weakly bound state. To separate the effects of two different RF signals from the effects of the positive ion, the model presented in this paper does not include the ion and the Coulomb interaction between the ion and the electron is not assumed. The motion of the electron is considered only due to the external trapping RF fields. The effects of the ion will be studied in a forthcoming study.

Using a purely classical approach, we formulate the equations of motion for the electron in the time-dependent trap potentials and cast them in the Hill / Floquet form. Stability and dynamics are then analysed in terms of the dimensionless q-parameters, which encode the applied voltages, drive frequencies, trap geometry, and the particle’s charge-to-mass ratio. Because the governing differential equations have the same structure for any charged species, the results for an electron can be translated to an ion by rescaling the q-values. Practically, this lets us (i) map stability regions in the experimentally accessible parameter space and (ii) identify, by rescaling, the parameter lines where the electron- and ion-stability regions overlap — i.e., the sets of trap voltages and frequencies that would allow both species to be confined under the same trap settings (note that this overlap criterion does not include Coulomb coupling).

Beyond stability maps, we characterise the electron’s classical motion via a Floquet decomposition. A lowest-order analytical truncation (secular term plus principal micromotion sidebands) captures the main spectral features — the slow secular oscillation and distinct low- and high-frequency micromotion components — while full numerical Floquet solutions identify where higher harmonics become important. These combined stability and dynamical results both specify the ranges of voltages/frequencies that permit confinement and quantify the frequency content and micromotion amplitudes relevant for applications such as RF sensing. Any analysis of coherent ion–electron dynamics would require including the Coulomb interaction and is therefore left for future work.

\section{Equations of Motion}
\label{sec:modelparams}
A Paul trap generates a dynamic potential in all three spatial directions and is capable of confining both positively and negatively charged particles. However, due to the substantial disparity in charge-to-mass ratios between light and heavy particles—such as electrons and ions—a single-frequency time-dependent potential cannot effectively trap both species simultaneously. To address this limitation, a dual-frequency Paul trap is employed \cite{Foot2018}.

We consider two distinct rf-voltages applied to electrodes in a geometry that generates an electric potential~\cite{major2005charged,Leibfried}
\begin{equation}
  \label{eq-potential}
  V(\mathbf{r},t) = \left( V_I \cos(\Omega_It) + V_e \cos(\Omega_et)\right)\frac{x^2+y^2-2z^2}{2r_0^2},
\end{equation}
where $r_0$ denotes the geometric scale of the trap. The rf-voltage with a lower angular frequency $\Omega_I$ and an amplitude $V_I$ is used to confine the \catwop ion, while the signal with a higher angular frequency $\Omega_e$ and a voltage amplitude $V_e$ is intended to trap a significantly lighter electron.

We analyze the dynamics of electron in the potential~\eqref{eq-potential} by means of the classical equation of motion
\begin{equation}
  \label{eq-eqom}
  \ddot{\mathbf{r}}_e = -Q_e\nabla V(\mathbf{r},t),
\end{equation}
where $Q_e$ is the charge-to-mass ratio of the electron. \eqab\eqref{eq-eqom} is decoupled in all spatial coordinates $x$, $y$, and $z$. While the discussion below is restricted to motion along the $x$-axis, the identical treatment applies to the remaining degrees of freedom. The $x$-component of the equation of motion yields
\begin{equation}
  \label{eq-eqomx}
  \ddot{x}(t) + \frac{Q_e}{r_0^2} \left(V_I\cos\left(\Omega_It\right)+V_e\cos\left(\Omega_et\right)\right)x(t)=0.
\end{equation}

Introducing the dimensionless time coordinate $\xi = \Omega_It/2$, the frequency ratio $\xi_r = \Omega_e/\Omega_I$, and the normalized trap parameters
\begin{equation}
  \label{eq-qparms}
  q_{I} = \frac{2 Q_e V_I}{r_0^2\Omega_I^2}, \qquad
  q_{e} = \frac{2 Q_e V_e}{r_0^2 \Omega_I^2},
\end{equation}
yields the transformation of \eqab\eqref{eq-eqomx} to the canonical form of a Hill differential equation \cite{Foot2018,McLachlan1947}
\begin{equation}
  \label{eq-Hill}
  \frac{d^2 x}{d \xi^2} - \left[2q_{I}\cos(2\xi)+2 q_{e} \cos(2\xi \xi_r) \right] x(\xi) = 0.
\end{equation}
Introduction of the parameters $q_{e,I}$ and $\xi_r$ in \eqab\eqref{eq-qparms} effectively reduces the number of parameters that determine the trajectories and, consequently, their stability. This transformation is a straightforward generalization of the standard approach to single-frequency rf-traps~\cite{Leibfried, major2005charged}.

\section{Conditions of Stability}
The form of \eqab\eqref{eq-Hill} is suitable for stability analysis of its solutions with respect to the values of $q_{e,I}$ and $\xi_r$ by means of the Floquet theory \cite{McLachlan1947}. For selected and fixed values of $q_{e,I}$ and $\xi_r$, this approach allows for expressing the trajectory as a Fourier-series-like linear combination of periodic functions. Their frequencies do not depend on the initial conditions and can be calculated without explicitly integrating the individual trajectories. When such a purely periodic expansion with only real frequencies is possible, the corresponding values of the parameters $q_{e,I}$ and $\xi_r$ are said to yield a stable regime of the trap, since the particle remains at a finite distance from the centre indefinitely. This mathematical approach is briefly reviewed in this section.

Assuming that $k\equiv\xi_r\in\mathbb{N}$, \eqab\eqref{eq-Hill} becomes a special case of the Hill's equation
\begin{equation}
  \label{eq_hillgen}
  \frac{d^2x}{d\xi^2}+\left(\theta_0+2\sum_{r=1}^\infty\theta_{2r}\cos(2r\xi)\right)x=0,
\end{equation}
where $\theta_0=0$ and the only non-zero coefficients $\theta_r$ are $\theta_2=q_I$, and $\theta_{2k} = q_e$.

Stable electron trajectories can also exist in situations where the frequency of the electron trap is not an integer multiple of the ion-trap frequency ($k\notin\mathbb{N}$). However, the method of stability analysis presented here is not applicable. Therefore, in the rest of this article, the discussion is restricted to the integer values of $\xi_r$. Since the frequencies of the trap are separated by several orders of magnitude, this technical limitation does not have any practical negative experimental implications.

According to Floquet's theorem, the stable solutions of \eqab\eqref{eq_hillgen} can be expressed as~\cite{Leibfried, McLachlan1947}
\begin{eqnarray}
  \label{eq_general solution}
  x(\xi)=Ae^{\mu_x \xi} \sum_{n=-\infty}^{\infty} C_{2n} e^{2in\xi}\nonumber\\
  +B e^{-\mu_x \xi} \sum_{n=-\infty}^{\infty} C_{2n} e^{-2in\xi},
\end{eqnarray} 
where $\mu_x$ is the characteristic exponent and $C_{2n}$ are the coefficients, both of which depend on $q_{e,I}$ and $k$. The solution is stable as long as the value of $\mu_x$ is purely imaginary.

Provided that $\theta_{2m}=\theta_{-2m}$ \cite{McLachlan1947,Leibfried}, substitution of the solution \eqref{eq_general solution} into \eqab\eqref{eq_hillgen} yields the recurrence relation
\begin{eqnarray}
  \label{eq_requrence_main}
  (\mu_x+2ni)^2 C_{2n}+q_I\left(C_{2n+2}+C_{2n-2}\right)\nonumber\\
  +q_e\left(C_{2n+2k}+C_{2n-2k}\right)=0.
\end{eqnarray}
Dividing throughout by 
\begin{equation}
    \label{eq-phicoeffs}
    \phi_{2r}\equiv(\mu_x +2ri)^2
\end{equation}
yields the set of homogeneous algebraic linear equations given by the infinite matrix
\begin{widetext}
  \begin{equation}
    \label{eq-linsys}
    \underline{M}(i\mu_x)=
    \begin{pmatrix}
      \multirow{5}{*}{\vdots} & \multicolumn{13}{c}{\dotsc} & \multirow{5}{*}{\vdots}\\
      & q_e/\phi_{-2} & 0 & 0 & \dotsc & q_I/\phi_{-2}  & 1               & q_I/\phi_{-2}  & 0               & 0         & \dotsc             & q_e/\phi_{-2}& 0             & 0           & \\
      & 0             & q_e/\phi_0 & 0 & \dotsc & 0 & q_I/\phi_0 & 1              & q_I/\phi_0 &  0        & \dotsc             & 0             & q_e/\phi_0 & 0           & \\
      & 0             & 0             & q_e/\phi_2 & \dotsc & 0               & 0 & q_I/\phi_2 & 1               & q_I/\phi_2 & \dotsc        & 0             & 0              & q_e/\phi_2 & \\
      & \multicolumn{13}{c}{\dotsc} & \\
    \end{pmatrix},
  \end{equation}
\end{widetext}
that is tri-diagonal with the units on the main diagonal and with two additional diagonals shifted $2k$ columns to the left and to the right from the main diagonal. It explicitly depends on $\mu_x$ via the coefficients $\phi_{2r}$ defined by \eqab\eqref{eq-phicoeffs}.

The solutions of the equation of motion \eqref{eq-Hill} are finite at all times when $\mu_x$ attains such a purely imaginary value that the system of linear equations \eqref{eq-linsys} has a solution. This can be expressed by the equation
\begin{equation}
  \label{eq-determinant}
  \Delta(i\mu_x)\equiv\det\left\{\underline{M}(i\mu_x)\right\}=0.
\end{equation}
When such a solution exists for given values of $q_{e,I}$ and $k$, the trap is in stable trapping mode. Therefore, the frequency spectrum of the trap, entirely determined by the value of $\mu_x$ (see \eqab\eqref{eq_general solution}), can be calculated without employing the initial conditions. Thi is an intrinsic feature of Floquet systems.

The determinant $\Delta(i\mu_x)$ is an even function of $\mu_x$ and its values are real as long as $q\in\mathbb{R}$. Additionally, $\Delta(i\mu_x+2i)=\Delta(i\mu_x)$, therefore it is sufficient to characterize it in the strip $0 \leq \text{Im}\{\mu_x\} \leq 1$. The only singularities of $\Delta(i\mu_x)$ are simple poles at $\mu_x=\pm 2ri$.

Adopting the method presented by \citet{McLachlan1947} (see also the Appendix in \cite{Trypogeorgos2016}), the stability condition \eqref{eq-determinant} can be solved using the analytical properties of the function $\chi(i\mu_x) = 1 / \left[\cos(i\pi\mu_x)-1\right]$ that shares the simple poles and periodicity with $\Delta(i\mu_x)$, reinforcing its structured analytic behavior. Consequently, the function $\phi(i\mu_x) = \Delta(i\mu_x)-C\chi(i\mu_x)$ does not have any singularities if $C$ is chosen appropriately. Since $\phi(i\mu_x)$ is a bounded entire function, it must be constant due to the Liouville's theorem. The value of the constant $C$ can be determined from the behavior as $|\mu_x| \to \infty$, where all the off-diagonal terms of matrix $\underline{M}(i\mu_x)$ vanish and
\begin{equation}
  \lim_{|\mu_x|\to\infty}\Delta(i\mu_x)=1.
\end{equation}
Since $\chi(i\mu_x)$ asymptotically vanishes, $\phi(i\mu_x)\equiv1$. This yields
\begin{equation}
  \label{eq_C_val_1}
  C=\frac{\Delta(i\mu_x)-1}{\chi(i\mu_x)}.   
\end{equation}
The value of $C$ can be easily evaluated for $\mu_x=i/2$ as $C=\left[1-\Delta(1/2)\right]$. Substitution into \eqab\eqref{eq_C_val_1} then yields the condition for the stable solution of the equation of motion as
\begin{equation}
  \label{eq_root_mu}
  \cos(i\pi\mu_x)=\Delta(1/2).
 \end{equation}
As $\mu_x$ is required to be purely imaginary, this condition can be expressed as
\begin{equation}
  \label{eq-stabcond}
  \left|\Delta(1/2)\right|\leq 1.
\end{equation}

As a result, the only numerical part of the stability study is the calculation of $\Delta(1/2)$ for a relevant range of $q_{e,I}$, $k$, and evaluating, whether its absolute value is less than 1. This is numerically very tractable task.

\section{Regions of Stability}
Using \eqab\eqref{eq-stabcond}, we explored the stability of the dual-frequency trap for a range of voltage amplitudes of the applied rf signals. In this analysis, the frequencies of the driving signals have been set at a fixed values $\Omega_I=2\pi\times25\thinspace$MHz and $\Omega_e=2\pi\times2\thinspace$GHz ($k = 80$).

This choice is motivated by the constraints of the experimental setup designed to co-trap the electrons and ions. Since the construction of an electron trap is predominantly tied to its resonant characteristic at frequency $\Omega_e$\cite{osada_feasibility_2022,Matthiesen20,kumar_3d-printed_2025}, it is necessary to choose its value already in the design stage. On the other hand, variation of $\Omega_I$ is comparatively less challenging, as the resonant circuitry is typically not embodied in the trap itself. Nevertheless, the trap is more tolerant to changing the voltage amplitudes $V_I$ and $V_e$ without changing its design. Therefore, the changes of $q_{e,I}$ within the studied range below are achieved only by changing the corresponding voltages $V_{e,I}$ (see \eqab\eqref{eq-qparms}). The assumed value of $r_0=\SI{1}{mm}$ has been obtained from the technical design of the trap described in the work by \citet{kumar_3d-printed_2025}.

\begin{figure*}
    \includegraphics[width=\textwidth]{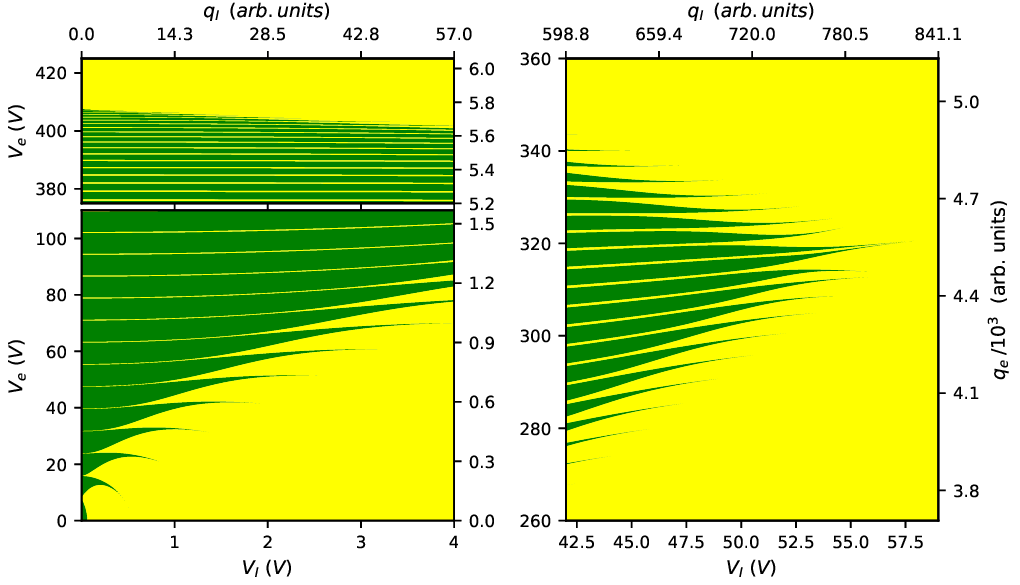}
   \caption{Stability diagram for $k = 80$, $\Omega_I=2\pi\times\SI{25}{MHz}$ and $\Omega_e=2\pi\times\SI{2}{GHz}$. The green areas are stable where the condition \eqref{eq-stabcond} is satisfied, the yellow regions are unstable. The parameters $q_I$ and $q_e$ on the top and right axes are mapped to the voltage amplitudes $V_I$ and $V_e$ on the bottom and left axes using \eqab\eqref{eq-qparms}, respectively. The stability landscape is presented in three panels. The bottom subplot reveals the emergence of a flame-shaped stability region beginning at \( q_e = 0 \), including embedded instability bands. The top subplot continues along the same \( q_I \) axis, extending to larger \( q_e \) values and illustrating the upper boundary of stability, beyond which all motion becomes unstable. The second column focuses on the regime of large \( q_I \), where stable trajectories exist only within a narrowing band of \( q_e \); eventually, stability is lost entirely beyond this range. Across all panels, the stability regions exhibit a bounded, flame-like structure—broadening with increasing \( q_I \) before tapering at the top.}
   \label{q_stability}
\end{figure*}

The stability diagram resulting from this parameter sweep is presented in \figab\ref{q_stability}, broken up into three panels.
It maps the regions in the $(V_I, V_e)$-parameter space where the characteristic exponent $\mu_x$ attains purely imaginary values, indicating stable trajectories (corresponding values of the parameters $q_{e,I}$ are presented on the secondary axes).

The presented structure of the stability region is notably flame-shaped. For each value of $q_I$, a minimum threshold of $q_e$ is required to achieve stability. This threshold gradually increases  with $q_I$. Simultaneously, the stability region is also bounded from above: beyond certain value of $q_e$, the system becomes unstable regardless of $q_I$. Furthermore, the range of stable $q_e$ for a fixed value of $q_I$ shrinks with increasing $q_I$. As a result, the stability region tapers off at high $q_I$, forming a narrowing flame-like structure in the $(q_I, q_e)$ plane. Within the stable region, narrow bands of instability appear periodically along the $q_e$-axis, consistent with the formation of the parametric resonances predicted by Floquet theory. These unstable strips divide the stable region into multiple disconnected lobes that appear in a regular pattern, and eventually vanish as the system approaches the edge of the stability zone.

These results are qualitatively consistent with those published by \citet{Trypogeorgos2016} for lower values of $k$. The stable regions exhibit similar structure affected by the parametric resonances. However, the size of these regions on the scale of the parameters $q_{e,I}$ is orders-of-magnitude different, due to very different masses considered by \citet{Trypogeorgos2016} (atomic or molecular ion) and in the present study (electron). Another source of a potential quantitative differences can be the damping of the system that is neglected here and considered in the reference \cite{Trypogeorgos2016}.

%\begin{figure}
%    \centering
%    \includegraphics[width=1\linewidth]{end_point.eps}
%    \caption{
%Stability regions in the \( (q_L, q_H) \) plane near the critical threshold for two different frequency ratios: (left) \( k = 95 \), (right) \( k = 110 \), with \( k = \Omega_H / \Omega_L \). Each plot shows a zoomed-in %view around the critical value of \( q_L \) beyond which no stable solutions exist, regardless of \( q_H \). The shaded regions represent values of \( (q_L, q_H) \) where the particle remains stably trapped.
%As \( k \) increases, the critical value of \( q_L \) shifts to larger values, indicating an expansion of the accessible stability region. This behavior suggests that increasing the frequency separation between the low- %and high-frequency drives enhances trap robustness and supports stable confinement over a wider range of parameters—an important feature for dual-species trapping scenarios.
%}
%    \label{fig:critical_qL}
%\end{figure}

The stability diagram becomes particularly relevant in the context of hybrid trapping systems, such as the simultaneous confinement of an electron and an ion. It can be unfolded by introduction of an additional set of parameters $q^I_{e,I}$ for the ion inside the two-frequency trap. They are defined as for \eqab\eqref{eq-qparms} with the charge-to-mass ratios of the ion $Q_I$ instead of those for the electron. The relation to the electron $q$-parameters is
\begin{equation}
    q^I_{e,I} = \left(\frac{Q_I}{Q_e}\right) q_{e,I},
\end{equation}
under identical applied voltages and trap geometry. Since $m_I \gg m_e$, the trap parameters of the electron are several orders of magnitude larger than those of the ion. For \catwop the ratio $Q_I/Q_e\approx2.74\times10^{-5}$. Therefore, when the voltages of the two-frequency trap are set within the region where the electron trapping is stable, the corresponding values of $(q^I_e,q^I_I)$ appear in the lowest stable lobe of the ion stability diagram.

The stability of the two-frequency trap, as discussed above, has been mapped throughout the whole manifold of $(V_e,V_I)$. However, only a section of this space is practically relevant from the perspective of the depth of the trap \cite{major2005charged}. In the experiment designed with the results of this study, at least in its initial stage, the electrons are intended to be trapped at room temperature. Therefore, the depth of the electron trap $\sim300$\thinspace K is necessary. This sets the lower limit for $V_e\gtrsim10$\thinspace V. Possible enclosing of the trap in a cryostat in the future could eventually allow for even lower voltage. The technical possibilities of the rf-signal source limit the upper limit to $V_e\lesssim120$\thinspace V.

Similarly, the lower bound of $V_I$ is determined by the depth of the ion trap. Since the ions will be optically cooled in the designed experiment down to the temperature $\sim1$\thinspace K, the corresponding trap depth can be achieved by $V_I\gtrsim1.5$\thinspace V. The upper limit is determined by the boundaries of the stability region in \figab\ref{q_stability}.

The lower boundaries of $V_{e,I}$ estimated above are only approximate within the given order of magnitude. The accurate values depend on the radial or axial direction inside the trap and on the specific geometry of the electrodes.

\section{Frequency analysis}
Once the condition of stability \eqref{eq-stabcond} is satisfied for a particular fixed set of voltages $V_{e,I}$ and angular frequencies $\Omega_{e,I}$ of driving signals, it is natural to investigate the frequency spectrum of the trapped electron. It is important in the context of the future quantum mechanical study of this system, as the frequency spectrum is related to the energy levels of the motional states of the trapped electron.
\subsection{Numerical Floquet Spectrum and Mode Structure}
The value of $\mu_x$ can be evaluated using \eqab\eqref{eq_root_mu}. Then the homogeneous system of linear algebraic equations $\underline{M}(i\mu_x)=0$ (see \eqab\eqref{eq-linsys}) can be numerically solved and the coefficients $C_{2n}$ defined by \eqab\eqref{eq_general solution} can be obtained.  Although their values can only be specified relative to each other without specifying the initial conditions (constants $A$ and $B$ in \eqab\eqref{eq_general solution}), it is sufficient to determine the relative contributions of the corresponding harmonic frequencies to the trajectories of the trapped electron.

\begin{figure}[htb]
  \includegraphics{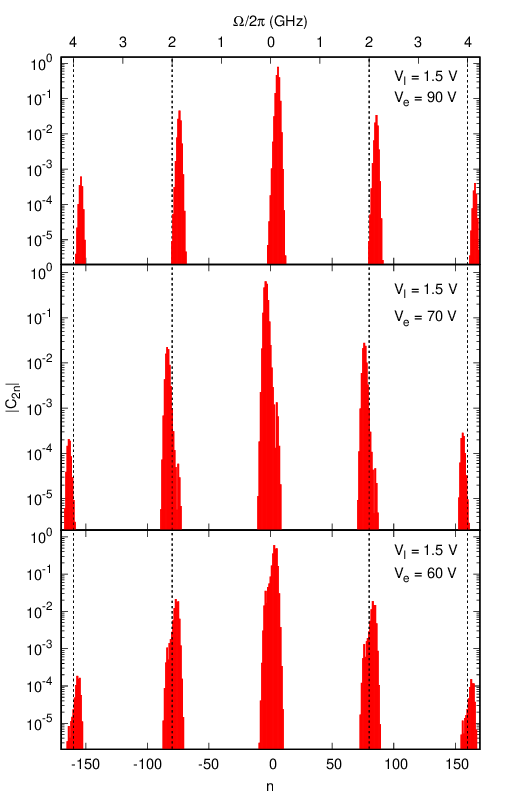}
  \caption{Expansion coefficients $C_{2n}$ defined by \eqab\eqref{eq_general solution} that determine the relative contribution of the harmonic modes to the motion of the electron in the two-frequency trap. Spectra for different voltage amplitudes $V_{e,I}$ of the driving rf signals are presented in the individual panels. Width of a single column corresponds to 25\thinspace MHz -- the driving frequency of the ion trap. The vertical dashed lines denote the integer multiples of the driving frequency of the electron trap (2\thinspace GHz). The bottom axis is the value of the index $n$ of the harmonic mode, the corresponding frequency $f=n\times 25\thinspace$MHz is on the top axis.}
  \label{fig:freqspec}
\end{figure}

The dominant part of the spectral profile of the two-frequency electron trap is presented in \figab\ref{fig:freqspec}.
It shows that several frequency modes with very low values of $n$ predominate the motion of the electron. A mode with index $n_0$ and angular frequency $\Omega_0=(-i\mu_x+2n_0)\Omega_I/2$ represents the dominant contribution followed by neighboring modes $n_0\pm1$. This triplet can be loosely interpreted as a ``canonical'' motion and a motion induced by the ion trapping rf signal. Although its character is similar to the micro motion known in the field of ion trapping, its amplitude is of the same or similar order as the canonical mode $n_0$. The higher modes further contribute to the motion of the trapped electron with decreasing amplitudes.
\begin{table}
    \caption{Index of the dominant mode $n_0$ and the secular frequency $\Omega_0/2\pi$ for the same values of the voltage amplitudes $V_{e,I}$ as in \figab\ref{fig:freqspec} and \figab\ref{fig:locfreqspec}. The last column shows the secular frequencies for $V_I=0$\thinspace V.}
    \label{tab:secfreqs}
    \begin{ruledtabular}
        \begin{tabular}{cccdd}
            $V_I$ (V) & $V_e$ (V) & $n_0$ & \multicolumn{1}{c}{$\Omega_0/2\pi$ (MHz)} & \multicolumn{1}{c}{$\Omega^{e}_0/2\pi$ (MHz)}\\
            \hline
            \multirow{3}{*}{1.5} & 60 &  3 & 84.27 & 105.16\\
                                 & 70 & -4 & 91.53 & 89.21\\
                                 & 90 &  6 & 158.13 & 157.10\\
            \hline
            \multirow{3}{*}{1} & 70  & -4 & 90.19 & 89.21\\
                               & 100 & -6 & 141.25 & 140.92\\
                               & 130 & -8 & 191.87 & 191.71\\
        \end{tabular}
    \end{ruledtabular}
\end{table}
The secular angular frequency $\Omega_0$ increases with $V_e$ (see \tabab\ref{tab:secfreqs}), reflecting a similar well known behavior in the trapping of the ions.

A particularly straightforward way to assess the influence of the ion trap on the trapping of the electron is a comparison of $\Omega_0$ with the secular frequency $\Omega_0^{e}$ obtained for $V_I=0$\thinspace V. It is shown in \tabab\ref{tab:secfreqs}. These two values approach each other with increasing $V_e$ and decreasing $V_I$. This demonstrates the possibility of such a combination of the voltage amplitudes where the interaction of both traps is maintained low, and simultaneously their depths are sufficient to confine the electron-ion pair.

As can be seen in \figab\ref{fig:freqspec}, the next dominant band of modes is close to the driving frequency of the electron trapping rf-signal. The dominant mode in this region is $n_0\pm k$ and its amplitude is one or two orders of magnitude smaller than the one of the canonical mode $n_0$. This can be interpreted as a ``micro motion'' due to the electron trapping rf signal. Similar peaks appear in the frequency spectrum at higher integer multiples of the electron driving frequency (modes $n_0\pm2k$, $n_0\pm3k$,\dots), each with a peak approximately two orders of magnitude below the peak of the previous band.

\begin{figure}[htb]
  \includegraphics{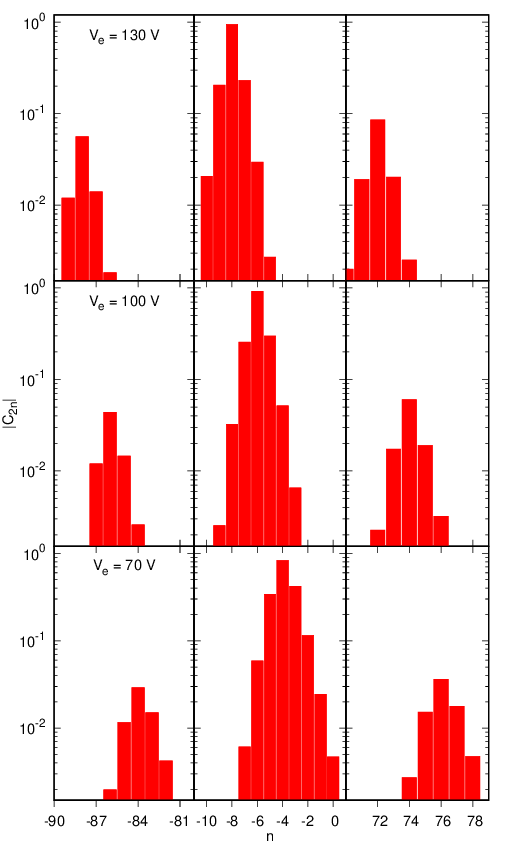}
  \caption{Expansion coefficients $C_{2n}$ defined by \eqab\eqref{eq_general solution}. Spectra for different voltage amplitudes $V_e$ and for a fixed value of $V_I=1$\thinspace V of the rf driving signals are presented in the individual panels. Width of a single column corresponds to 25\thinspace MHz -- the driving frequency of the ion trap. The horizontal axis is the index $n$ of the harmonic mode.}
  \label{fig:locfreqspec}
\end{figure}

Although only a single value of $V_I=1.5$\thinspace V is presented in \figab\ref{fig:freqspec}, the features of the frequency spectrum described above are consistent across the entire domain of $(V_e, V_I)$ where the trap is stable. All of the dominant bands mentioned above are narrow and clearly separated. Nevertheless, they generally consist of multiple frequency modes. The details of the most important bands are presented in \figab\ref{fig:locfreqspec} for $V_I=1$\thinspace V.
It shows that the contribution of the dominant modes $n_0$, $n_0\pm1$, $n_0\pm k$ increases relatively to the other modes with increasing $V_e$. As the effect of the electron trap increases relative to the effect of the ion trap, the frequency spectrum approaches the structure known from the single-frequency ion traps, where all modes, except the canonical motion and micro-motion, can be neglected. However, while there is usually only a single mode corresponding to the micro-motion in the case of the ions trapped by a single frequency, this motion has two components in the two-frequency electron trapping: one is due to the ion trapping signal, another is due to the electron trap.

\begin{figure}[htb]
  \centering
  \includegraphics[width=\linewidth]{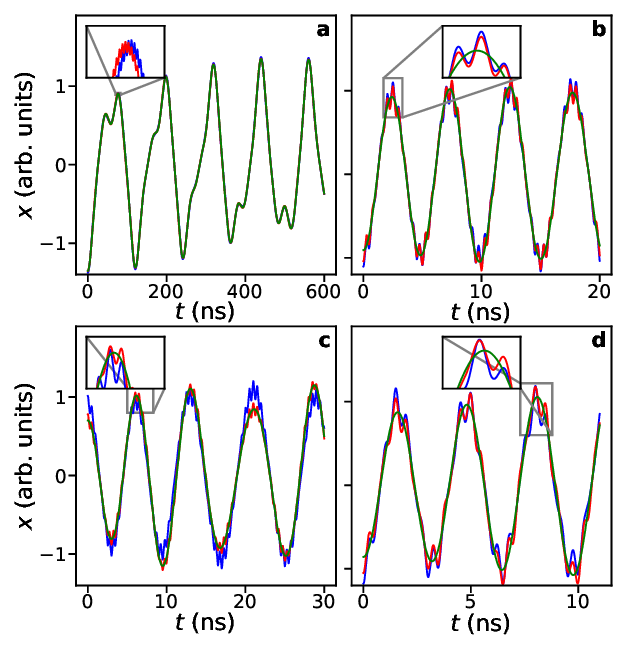} % Replace with your figure file
  \caption{Comparison of electron trajectories reconstructed using specific subsets of Fourier components for drive frequencies 
  (a) Small $q$-parameter regime ($q_I = 0.58$, $q_e = 44$). 
  (b-d) Large $q$-parameter regime ($q_I = 14.6$, $q_e = 1750$,$q_I = 22$, $q_e = 1300$,$q_I = 28$, $q_e = 2600$). 
  The blue solid line represents the full numerical solution (modes $n = -400$ to $400$), the green line shows the canonical motion triplet $(n_0, n_0 \pm 1)$, and the red line incorporates both the canonical motion and the high-frequency electron-trap sidebands $(n_0 \pm k)$.}
  \label{fig:qualitative_structure}
\end{figure}

\subsection{Lowest-Order Analytical Approximation}

To gain analytical insight into the physical mechanisms governing these spectral features, much of the trajectory behavior can be captured using a lowest-order analytical approximation. While exact trajectory construction requires solving a large system of coupled algebraic equations, this analytical model isolates the dominant frequency components and provides an intuitive framework for interpreting the trajectory structure. {Since the results presented in this subsection have broader validity, they are expressed in terms of the dimensionless parameters $q_e$ and $q_I$ defined in \eqab\eqref{eq-qparms}.} 

In the preceding numerical Floquet analysis, the characteristic exponent \(\mu_x\) is represented within a principal interval, while the dominant contribution to the motion may correspond to a nonzero Fourier-mode index \(n_0\). For the analytical approximation developed here, we adopt an equivalent Floquet representation in which this dominant mode is shifted to the central index \(n=0\). This is achieved by introducing the shifted characteristic exponent \[
\mu_{\mathrm{dom}}=\mu_x+2in_0.
\]
For stable solutions, we write \(\mu_{\mathrm{dom}}=i\tilde{\mu}\), where \(\tilde{\mu}\) is real. Thus, \(\tilde{\mu}=-i\mu_x+2n_0\) directly determines the dominant secular frequency, \(\Omega_0=\tilde{\mu}\Omega_I/2\). The corresponding Fourier coefficients are reindexed consistently.

The starting point is the general solution in \eqab\eqref{eq_general solution}, expressed as a Floquet-Fourier series in harmonics of the trap drive frequencies. As demonstrated in Fig. \ref{fig:qualitative_structure},
%the dominant features of the numerical trajectories can be reproduced by retaining only a small subset of Fourier components. While the three-component description captures the canonical motion and its low-frequency micromotion, including the high-frequency electron-trap sidebands provides a more complete representation of the trajectory. 
in the small $q_I$ and $q_e$ regime (case (a)), the trajectory closely resembles motion in a standard single-frequency Paul trap, characterized by a slow secular oscillation modulated by low-frequency micromotion. Additionally, high-frequency micromotion driven by the electron-trapping RF signal appears as a subtle ripple superimposed on the motion.
%, which the canonical approximation $(n_0, n_0 \pm 1)$ cannot capture.

Conversely, in the large $q_I$ and $q_e$ regime (cases (b-d)), where $\tilde{\mu} > 15$, the sideband frequencies $(\tilde{\mu} \pm 2)\Omega_I/2$ lie close to the secular frequency and no longer produce distinct, well-separated micromotion. Instead, they introduce a slight frequency shift and modulation into the secular trajectory itself. At the same time, high-frequency micromotion from the electron-trapping RF signal remains clearly defined, with an amplitude scaled up by the larger $q_e$ value.
%In this regime, the canonical approximation fails to capture this pronounced micromotion, highlighting the necessity of including the $n_0 \pm k$ modes.

Motivated by the trajectories in Fig. \ref{fig:qualitative_structure}, we now formulate a lowest-order analytical approximation to the exact solution \eqref{eq_general solution} by retaining only the dominant secular term \(d_0\) and the primary sidebands \(d_{\pm 2}\) and \(d_{\pm 2\xi_r}\). This five-component truncation generalizes the familiar three-component description of motion in a single-frequency Paul trap \cite{Leibfried}. The two-frequency system is represented by including the additional sidebands associated with the second driving field.

The necessity of these additional sidebands is apparent from the three-component trajectories presented in \figab\ref{fig:qualitative_structure} along with the accurate solutions \eqref{eq_general solution}. While the three-component description captures the canonical motion and its low-frequency micromotion, the high-frequency electron-trap sidebands are missing. Specifically, in the case (a), it does not resolve the subtle ripples superimposed on the motion by the high-frequency micromotion. Similarly, in the cases (b)-(d), the low-frequency approach omits the slight frequency shift and modulation of the low-frequency secular trajectory.

Physically, these five terms, necessary to capture all qualitative features of the trajectories, correspond to their distinct frequency components: The secular motion at frequency \(\tilde{\mu}\Omega_I/2\), sideband micromotion due to \((\tilde{\mu}\pm 2)\Omega_I/2\), and high-frequency micromotion originating from \((\tilde{\mu}\pm 2\xi_r)\Omega_I/2\). Assuming \(A=B=1\), the resulting approximate trajectory takes the form:

\begin{equation}\label{eq_appro_cos}
\begin{aligned}
x(t) \approx &\; \cos \left( \tilde{\mu} \frac{\Omega_{I} t}{2} \right) + d_{\pm2} \cos\left( \frac{(\tilde{\mu} \pm 2)\Omega_I t}{2} \right) \\
&+ d_{\pm2\xi_r} \cos\left( \frac{(\tilde{\mu} \pm 2\xi_r)\Omega_I t}{2} \right).
\end{aligned}
\end{equation}

In this approximation, when the system operates in a stable regime, the recurrence relation \eqab\eqref{eq_requrence_main} simplifies to a finite system from which we express:
\begin{equation}
d_{\pm2} \approx \frac{-q_I}{(\tilde{\mu} \pm 2)^2} d_0, \qquad d_{\pm2\xi_r} \approx \frac{-q_e}{(\tilde{\mu} \pm 2\xi_r)^2} d_0.
\end{equation}

Substituting these expressions back into the recurrence relation at $n = 0$ yields an implicit equation for the characteristic exponent $\mu_x$:
\begin{equation}
(\tilde{\mu})^2 - \frac{q_I^2}{(\tilde{\mu} \pm 2)^2} - \frac{q_e^2}{(\tilde{\mu} \pm 2\xi_r)^2} = 0.
\end{equation}
In the regime where $q_I$ is small and $\xi_r$ is large, this reduces to a simple closed-form approximation:
\begin{equation}\label{eq_mu_quadrature}
\tilde{\mu}_{\text{approx}} \approx \pm  \sqrt{ \frac{q_I^2}{2} + \frac{q_e^2}{2\xi_r^2} },
\end{equation}
which explicitly reveals how the secular frequency depends on both driving fields, showing that the high-frequency drive contribution is suppressed by $1/\xi_r$. (See Appendix \ref{app}). 

The approximation in Eq.~\eqref{eq_mu_quadrature} provides a explicit estimate of the shifted characteristic exponent associated with the dominant Fourier component. Unlike the numerically accurate Floquet solution, the five-component truncation does not automatically account for changes in the dominant Fourier-mode index across different branches of stability in the space $(q_e,\quad q_I)$. Consequently, its range of accuracy is limited.
This domain of applicability could be expanded by adding $n_0$ as an additional parameter into the model.

\begin{figure}[htb]
  \centering
  \includegraphics[width=\linewidth]{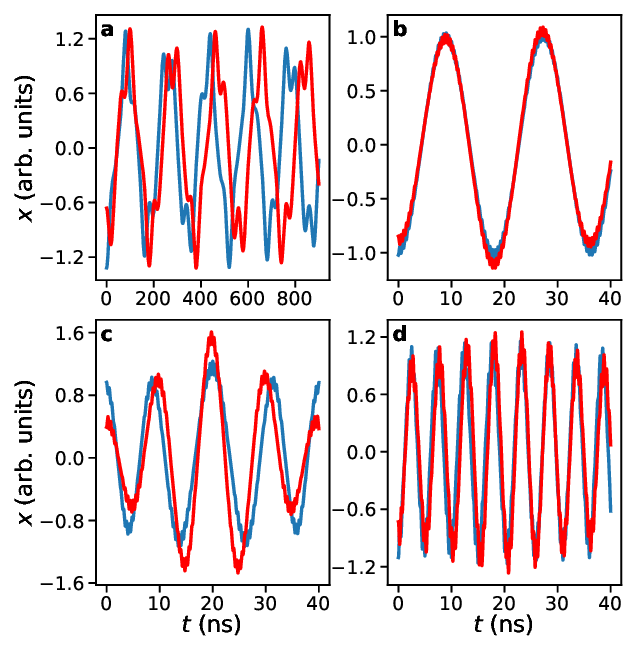}
  \caption{Comparison between the exact numerical solution of Eq.~(\ref{eq_general solution}) (blue line) and the lowest-order analytical approximation given in Eq.~(\ref{eq_appro_cos})(red line) for $A = B = 1$, with drive frequencies $\Omega_I = 2\pi \times \SI{25}{MHz}$ and $\Omega_e = 2\pi \times \SI{2}{GHz}$ ($k = 80$). The exact solution is computed using 801 Fourier components ($n = -400$ to $n = 400$), while the approximation retains only five terms: $d_0$, $d_{\pm2}$, and $d_{\pm2\xi_r}$. Each panel corresponds to a different pair of parameters $(q_I, q_e)$:
(a) $(q_I=0.587, q_e=11)$,
(b) $(q_I=0.587, q_e=500)$,
(c) $(q_I=14.62, q_e=916)$, and
(d) $(q_I=14.62, q_e=1750)$.}
  \label{fig:lowest_app}
\end{figure}

To evaluate the accuracy of this lowest-order approximation, we compare it with the exact numerical trajectory obtained by solving 801 coupled equations as defined in \eqab\eqref{eq-linsys}. Figure~\ref{fig:lowest_app} presents the approximate trajectory from Eq.~\eqref{eq_appro_cos} alongside the exact solution from \eqab\eqref{eq_general solution}, constructed using the full set of coefficients for $n = -400$ to $n = 400$ with $A = B = 1$.

Over a wide range of $q_e$ and $q_I$ values, the lowest-order approximation accurately captures the dominant structure of the particle's motion. However, in the narrow stable tongues the five-component truncation becomes insufficient. In these regions, the motion is governed by a larger number of significant Fourier components. Nevertheless, because these stable tongues are extremely narrow and experimentally susceptible to particle loss from minor voltage or frequency fluctuations, they are generally not targeted in practical experimental setups.

\begin{figure}
    \centering
    \includegraphics[width=0.98\linewidth]{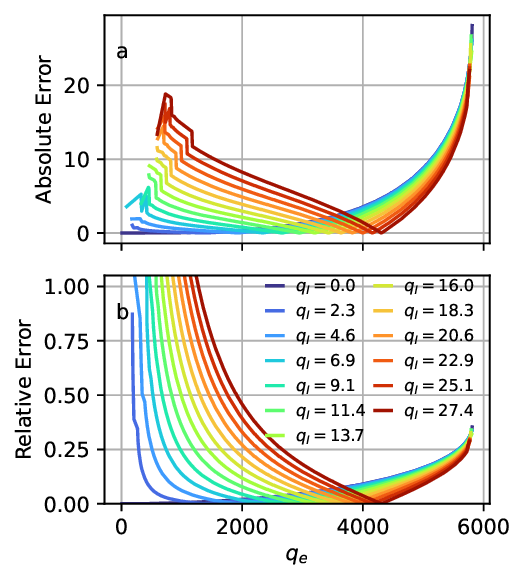}
    \caption{Comparison of the numerical characteristic exponent \(\mu_{\mathrm{num}}\) with the five-component approximation \(\tilde\mu_{\mathrm{approx}}\) from Eq.~(20) across the stable regions in the \((q_I,q_e)\) plane. (a) Absolute error, \(|\mu_{\mathrm{num}}+2in_0-i\tilde\mu_{\mathrm{approx}}|\). (b) Relative absolute error, \(|\mu_{\mathrm{num}}+2in_0-i\tilde \mu_{\mathrm{approx}}|/|\mu_{\mathrm{num}}+2in_0|\).}
    \label{fig:error}
\end{figure}

{  The characteristic exponent \(\mu_x\) is an important quantity in the five-component approximation because it determines the main frequency of the electron motion. To assess the accuracy of its prediction, we compare the approximate value \(i\tilde\mu_{\mathrm{approx}}\), obtained from \eqab\eqref{eq_mu_quadrature}, with the numerical value \(\mu_{\mathrm{num}}\) calculated from the full system of coupled equations \eqab\eqref{eq_requrence_main}. Figure~\ref{fig:error} shows the absolute error, \(|\mu_{\mathrm{num}}+2in_0- i\tilde\mu_{\mathrm{approx}}|\), in panel (a) and the relative absolute error, \(|\mu_{\mathrm{num}}+2in_0-i\mu_{\mathrm{approx}}|/|\mu_{\mathrm{num}}+2in_0|\), in panel (b), within the stable regions in the \((q_I,q_e)\) parameter space. For clarity, panel (b) displays only the regions where the relative error is less than one. For values of \(q_I\) beyond the first stability region shown in Fig. \ref{q_stability}, the absolute error initially exhibits relatively large fluctuations over the ranges of \(q_e\) corresponding to the stability tongues. As \(q_e\) increases, the absolute error then decreases approximately linearly, reaching a minimum close to zero at a threshold value of \(q_e\) for each \(q_I\). Beyond this threshold, the error increases rapidly, exhibiting an approximately exponential growth. However, as shown in panel (b), the relative error remains small even as \(q_e\) increases. This indicates that  despite the increasing absolute difference, the approximation can retain good relative accuracy over a substantial part of the stable parameter space. Nevertheless, the approximation is less reliable in the vicinity of stability-region boundaries. These results complement the trajectory comparisons presented above and clarify the range over which the five-component approximation provides a useful estimate of the characteristic exponent.}
\section{CONCLUSION}
We have presented a classical theoretical study of electron dynamics in a two-frequency Paul trap designed for a hybrid electron--$\text{Ca}^{2+}$ system. The primary purpose of this single-electron analysis is to establish a rigorous baseline reference case in the absence of the ion. By isolating the electron from ion--electron interactions, we have mapped its unperturbed stability boundaries and identified its characteristic frequency spectrum under the combined action of the two radio-frequency fields.

The space of the explored parameters of the system corresponds to the conditions in the experiment designed to study the Rydberg states of Ca$^+$. The selected quantities and their ranges reflect the anticipated experimental circumstances and this study confirms that a stable regime of the combined electron-ion trap can be achieved.

Our stability analysis confirms the characteristic flame-like structure previously reported for two-frequency Paul traps in the $(q_I, q_e)$ parameter space. Keeping the drive frequencies fixed at $\Omega_I = 2\pi \times 25\text{ MHz}$ and $\Omega_e = 2\pi \times 2\text{ GHz}$ ($k = 80$), we systematically varied the signal voltage amplitudes $(V_I, V_e)$ to identify the accessible stability regions and locate trap parameters that permit stable electron confinement alongside ion stability.

In addition to mapping stability, we evaluated the spectral content of the trapped electron's motion. A five-component analytical Floquet model retaining only $d_0$, $d_{\pm 2}$, and $d_{\pm 2k}$ was shown to accurately capture the exact numerical trajectory across a wide range of voltages. This model provides a clear physical picture of the electron motion as a slow secular oscillation modulated by two distinct micromotion sidebands originating from the low- and high-frequency RF drives.

%It is important to emphasize that this single-electron framework is not intended to describe the complete dynamics of the coupled electron--ion system. Instead, it provides the benchmark dynamics required for subsequent studies. When evaluating the full coupled electron--ion system in future work, comparing the resulting coupled motion against the reference secular and micromotion frequencies derived here will directly reveal and quantify the frequency shifts induced by the presence of the ion. This framework thus serves as a necessary foundation for designing and optimizing hybrid two-frequency Paul traps for precision sensing and quantum control.

The model presented in this study does not explicitly include the ion. Even though the dynamics of the electron-ion system is formally not complete, the results presented above are in remarkable correspondence with the preliminary findings of advanced calculations that also include the ion (beyond the scope of this paper). This suggests that the electron dynamics due to the two-frequency trap predominates the effects of the cation, at least in the dominant features in the frequency domain. The detailed non-linear interplay between the potential of the ion and the two-frequency trap, particularly in the structure of the trajectories, will be a subject of a forthcoming work. The study presented here establishes a benchmark for the mechanisms studied there.

\begin{acknowledgments}
This work is supported by the Czech Science Foundation (GA\v CR: GA24-10992S).
\end{acknowledgments}

%\bibliography{mainbib,referencesMH}% Produces the bibliography via BibTeX.

%
\bibliographystyle{apsrev4-2}

\bibliography{mainbib,xrefs}
\clearpage            % Ensures main text finishes on its own two-column page
\onecolumngrid

\appendix

\section{Lowest-Order Approximation} \label{app}

In the lowest-order approximation for a two-frequency Paul trap, only the Fourier components \( d_0 \), \( d_{\pm 2} \), and \( d_{\pm 2\xi_r} \) are retained, while all higher-order coefficients are neglected. The resulting expression for the solution becomes:
\begin{equation}
\begin{aligned}
    x(\xi) \approx &\; A\left(d_{\pm2} e^{i(\pm2+\tilde{\mu})\xi} + d_0 e^{i\tilde{\mu}\xi} + d_{\pm2\xi_r} e^{i(\pm2\xi_r+\tilde{\mu})\xi}\right) \\
    &+ B\left(d_{\pm2} e^{i(\pm2-\tilde{\mu})\xi} + d_0 e^{-i\tilde{\mu}\xi} + d_{\pm2\xi_r} e^{i(\pm2\xi_r-\tilde{\mu})\xi}\right)
\end{aligned}
\end{equation}

When \( a = 0 \), the coefficients \( d_{\pm2} \) and \( d_{\pm2\xi_r} \) are determined from the recurrence relation associated with stable solutions:
\begin{equation}\label{ap_requrence_main}
   (\tilde{\mu} + 2n)^2 d_{2n} + q_I(d_{2n+2} + d_{2n-2}) + q_e(d_{2n+2\xi_r} + d_{2n-2\xi_r}) = 0
\end{equation}
Evaluating this for \( n = 0, \pm1, \pm\xi_r \), we get:
\begin{align}
    &\tilde{\mu}^2 d_0 + q_I (d_{2} + d_{-2}) + q_e (d_{2\xi_r} + d_{-2\xi_r}) = 0 \\
    &(\tilde{\mu} \pm 2)^2 d_{\pm2} + q_I (d_{\pm 4} + d_0) + q_e (d_{\pm2 + 2\xi_r} + d_{\pm2 - 2\xi_r}) = 0 \\
    &(\tilde{\mu} \pm 2\xi_r)^2 d_{\pm2\xi_r} + q_I (d_{\pm2\xi_r + 2} + d_{\pm2\xi_r - 2}) + q_e (d_{\pm4\xi_r} + d_0) = 0
\end{align}

Assuming only \( d_0 \), \( d_{\pm2} \), and \( d_{\pm2\xi_r} \) are non-zero, we obtain:
\begin{equation}\label{ap_C_2_C_80}
    d_{\pm2} \approx -\frac{q_I}{(\tilde{\mu} \pm 2)^2} d_0, \qquad
    d_{\pm2\xi_r} \approx -\frac{q_e}{(\tilde{\mu} \pm 2\xi_r)^2} d_0
\end{equation}

Substituting these into the recurrence relation at \( n = 0 \) gives:
\begin{equation} \label{ap_root}
    \tilde{\mu}^2 - q_I^2 \left( \frac{1}{(\tilde{\mu} + 2)^2} + \frac{1}{(\tilde{\mu} - 2)^2} \right)
             - q_e^2 \left( \frac{1}{(\tilde{\mu} + 2\xi_r)^2} + \frac{1}{(\tilde{\mu} - 2\xi_r)^2} \right) = 0
\end{equation}

Alternatively, expressing in terms of scaled variables:
\begin{equation} \label{ap_root_scaled}
    \tilde{\mu}^2 - \frac{q_I^2}{4} \left( \frac{1}{(\tilde{\mu}/2 + 1)^2} + \frac{1}{(\tilde{\mu}/2 - 1)^2} \right)
             - \frac{q_e^2}{(2\xi_r)^2} \left( \frac{1}{(\tilde{\mu} / 2\xi_r + 1)^2} + \frac{1}{(\tilde{\mu} / 2\xi_r - 1)^2} \right) = 0
\end{equation}

Using Newton's generalized binomial theorem, we expand:
\begin{align}
    \frac{1}{\left(\frac{\tilde{\mu}}{2} \pm 1\right)^2} &= 1 \mp \tilde{\mu} + \frac{3}{4} \tilde{\mu}^2 \mp \tilde{\mu}^3 + \dots \\
    \frac{1}{\left(\frac{\tilde{\mu}}{2\xi_r} \pm 1\right)^2} &= 1 \mp \frac{\tilde{\mu}}{\xi_r} + \frac{3}{4} \left(\frac{\tilde{\mu}}{\xi_r}\right)^2 \mp \left(\frac{\tilde{\mu}}{\xi_r}\right)^3 + \dots
\end{align}

Combining the expansions:
\begin{align}
    \tilde{\mu}^2 \Bigg[1 - \frac{3}{8} q_I^2 - \frac{3}{8\xi_r^2} q_e^2 + \mathcal{O}(\tilde{\mu}^2) \Bigg]
    \approx \frac{q_I^2}{2} + \frac{q_e^2}{2\xi_r^2}
\end{align}

Retaining terms only up to \( \tilde{\mu}^2 \), we obtain the approximate relation:
\begin{equation}
    \tilde{\mu}^2 \approx \frac{q_I^2}{2} + \frac{q_e^2}{2\xi_r^2}
\end{equation}

This expression provides a compact approximation for the shifted characteristic exponent associated with the dominant Fourier component. It also shows that the contributions of the low- and high-frequency driving fields, represented by the parameters \(q_I\) and \(q_e\), respectively, combine in quadrature.

\end{document}